# COULOMB SUMS AND MODIFICATION OF NUCLEONS IN THE ATOMIC NUCLEUS

A. Yu. Buki

*National Science Center "Kharkov Institute of Physics and Technology", Kharkov, Ukraine*

Experimental values of Coulomb sums for q > 2 fm$^{-1}$ are considered. Assuming that a part of nucleons is modified in the nucleus, an approach is offered, in the framework of which the calculation is in agreement with the experimental data. It gives the nuclear matter density value, above which the nucleon changes its properties. The radius of a modified proton is estimated, the possibilities for a further investigation into its properties are discussed.

## Introduction

The cross section for electron scattering from the atomic nucleus at momentum transfers higher than 2 fm$^{-1}$ is determined by quasi-elastic scattering (QES) from nucleons. So, it might be expected that at the mentioned momentum transfers the cross section for scattering from the nucleus should be equal to the sum of cross sections for scattering from nucleons making up the nucleus. However, the experiments show that the expected equality of cross sections is not observed for most nuclei [1,2]. This problem has been comprehensively considered (e.g., see review [1] and papers [2,3]), but the approaches proposed fail to provide a satisfactory explanation of the totality of the available experimental data. One of the approaches applied for the analysis of QES is based on the hypothesis of nuclear nucleon modification. The present paper suggests the modification of only a part of nucleons in the nucleus. Conditions at which the nucleus is in the modified state are investigated.

## 1. Nomenclature and experimental data

1.1 The differential inelastic electron scattering cross section $d^2\sigma(\theta,E,E')$ can be described in terms of the longitudinal $R_L(q,\omega)$ and transverse $R_T(q,\omega)$ response functions [4]:

$$d^2\sigma(\theta,E,E') = \sigma_M(\theta,E)\,(G^p_E(Q^2))^2\,\{\lambda^2 R_L(q,\omega) + [\lambda/2 + \tan^2(\theta/2)]\,R_T(q,\omega)\}, \qquad (1)$$

where $\sigma_M(\theta,E)$ is the Mott cross section; $G^p_E(Q^2)$ is the electric form factor of a proton; $E$ is the initial electron energy; $E'$ is the energy of electron scattered through the angle $\theta$ ; $\omega$, $q$ and $Q$ are, respectively, the energy transfer, the three- and four-momentum transfers; $\lambda = Q^2/q^2$. The Coulomb sum (CS) normalized to the number of protons in the nucleus $Z$ is

$$S(q) = \int\limits_0^\infty \frac{R_L(q,\omega)}{Z\,\eta(Q^2)}\,d\omega, \qquad (2)$$

where the relativistic correction $\eta(Q^2)$, according to ref. [5], is given by $(1+Q^2/4M^2)/(1+Q^2/2M^2)$, and $M$ is the proton mass. As noted above, for $q \approx 2$ fm$^{-1}$ the electron-nucleus scattering cross section is determined by QES. Therefore, if it is assumed that the electromagnetic form factor of nuclear proton does not differ from the form factor of a free proton, then integral (2) should not depend on $q$ and $S(q) = 1$. We shall write $S(q \geq 2$ fm$^{-1})$ as $S'$.

1.2 The experimental $R_L(q,\omega)$ values for the nuclei $^{2,3}$H, $^{3,4}$He, $^{12}$C, $^{40,48}$Ca, $^{56}$Fe and $^{208}$Pb were obtained at the Saclay and Bates laboratories [2,6]. The measurements were carried out in the range $q = 1.5 \div 3$ fm$^{-1}$ up to the energy transfers $\omega_{max} = 76 \div 360$ MeV. As can be seen from Fig.1, the measurements cover nearly all the area of the QES peak. This provides integral (2) values up to $\omega_{max}$, so $S(q,\omega_{max}) \approx S(q)$. Fig.2 shows $S(q,\omega_{max})$ of the nuclei $^4$He, $^{48}$Ca, $^{208}$Pb as a function of $q$. This behavior of experimental CS is typical of all the nuclei under study. The results of measurements at $q > 2$ fm$^{-1}$ can be summarized as the following regularities:

I. Within the limits of experimental error, the CS at $q \geq 2$ fm$^{-1}$ is independent of the momentum transfer to within the experimental error. Such a behavior of $S(q)$ corresponds to QES.

II. For $^2$H, $^3$H and $^3$He, we have $S' \cong 1$, that is expected in the QES case. However, $S' < 1$ for all nuclei with $A \geq 4$, and the $D = 1 - S$ difference value exceeds the error of $S'$ measurement, $\Delta S'$.

III. The $D$ value increases with an increasing atomic number of nucleus.

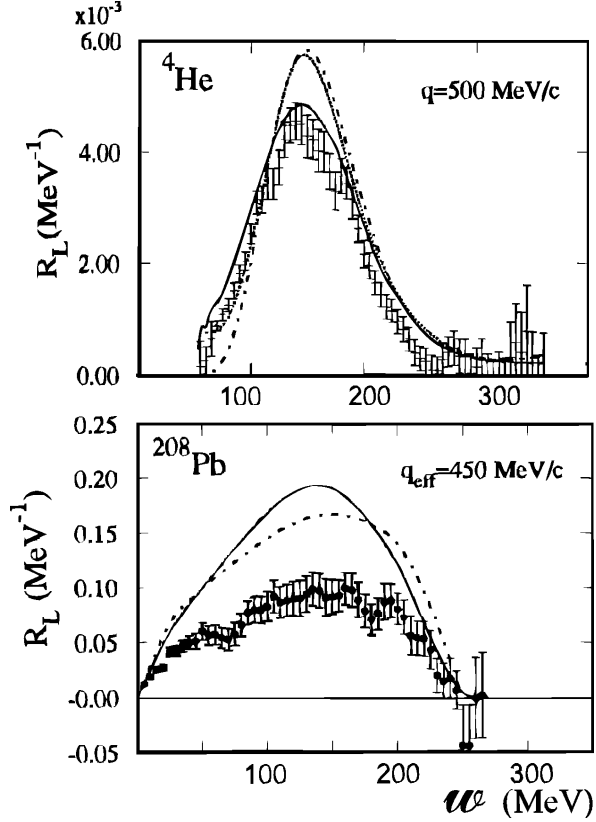

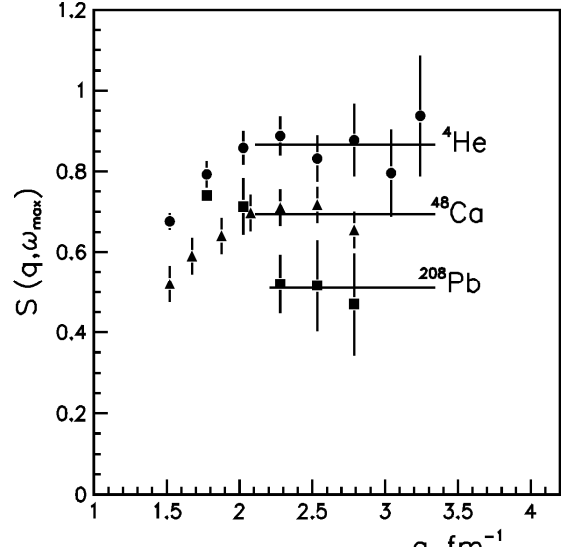

Fig.1. Experimental values and calculation of longitudinal response functions for the nuclei $^4$He, $^{208}$Pb of [2].

Fig.2. C.s. $S(q,\omega_{max})$ as a function of momentum transfer for the nuclei $^4$He, $^{48}$Ca, $^{208}$Pb of [2].

## 2. Analysis of experimental data

2.1 The CS value is determined by the interaction of electron with the nuclear charge. Therefore, we shall try to find the relationship between the CS and the charge density distribution in the nucleus, $\rho_e(r)$. In the case of $A<4$ ($S'=1$), almost all the volume of nucleus falls within the diffuse layer of charge distribution, whereas in heavier nuclei ($S'<1$), a significant part of charge is concentrated in the region, where $\rho_e(r)$ is close to its highest values in this nucleus, i.e., $\rho_e(r)/\max(\rho_e(r)) > 0.9$.

Let us assume that QES is mainly contributed by the external region of the nucleus, where the charge distribution has a diffuse form. It is obvious that in this case regularity (II) should hold true. And since the part of nuclear volume belonging to the diffuse layer decreases with increasing $A$, then the CS decreases, too, and this corresponds to regularity (III). For the spherical nucleus, this hypothesis is written as

$$S' = 1 - \frac{4\pi}{Z} \int_0^{r_o} \rho_e(r)\, r^2 dr \,, \qquad (3)$$

where $r_0$ is the radius of sphere separating the outer protons involved in QES from the inner protons which do not participate in the process. The integral in Eq.(3) determines the fraction of protons which do not contribute to QES, it is denoted above as D.

To go over to the quantitative study of the problem, let us average the $S(q,\omega_{max})$ values from ref. [2] in the range of $q = 2.5\pm0.3$ fm$^{-1}$ (for $^4$He, $q = 2.5\pm0.5$ fm$^{-1}$), and take into account the correction $T$ for extrapolation of $R_L$-functions to the region $\omega > \omega_{max}$: $S' = \overline{S}_L(q,\omega_{max}) + T$. With the given data, according to [2,6,7], we have $T = (0.06\pm0.03)\times S'$ for $4 \le A \le 56$ and $T = (0.10\pm0.05)\times S'$ for $^{208}$Pb. The $\overline{S}_L(q,\omega_{max})$ and $S'$ values thus obtained are listed in Tab.1.

Using Eq.(3) we determine the $r_0$ and $\rho_e(r_0)/\max(\rho_e(r))$ values (Table 2). The choice of the form of function $\rho_e(r_0)$ from the realistic models, as well as the experimental errors of parameters of this function little reflect on the $r_0$ value, and therefore, $\Delta r_0$ is determined by $\Delta S'$. It is seen from Table 2 that the $r_0$ and $\rho_e(r_0)/\max(\rho_e(r))$ values are consistent with the assumption about the role of nuclear diffuse layer in QES.

Table 1 $\bar{S}(q,\omega_{max}) \pm \Delta\bar{S}(q,\omega_{max})$ are the averaged experimental Coulomb sum values from [2]. $S' \pm \Delta S'$ are the same data extrapolated over $\omega_{max}$ to infinity.

|  | $^4$He | $^{12}$C | $^{40}$Ca | $^{48}$Ca | $^{56}$Fe | $^{208}$Pb |
|---|---|---|---|---|---|---|
| $\bar{S}(q,\omega_{max})$ | 0.862 | 0.792 | 0.602 | 0.694 | 0.705 | 0.510 |
| $\pm\Delta\bar{S}(q,\omega_{max})$ | 0.025 | 0.040 | 0.062 | 0.026 | 0.078 | 0.055 |
| $S'$ | 0.913 | 0.842 | 0.638 | 0.736 | 0.747 | 0.561 |
| $\pm\Delta S'$ | 0.051 | 0.064 | 0.080 | 0.047 | 0.100 | 0.081 |

Table 2. The values of radius of the sphere separating the outer nuclear protons which give, according to the hypothesis, the main contribution to the Coulomb sum (calculations by Eq.(3)). The models of charge distribution in the nucleus are designated as: $S$ (shell model); $F2$ and $F3$ (two- and three-parametric Fermi models, respectively). For the Fermi models given are the values of the parameters $c$ the radius of density half-falloff; $b$ diffusivity; $w$: density variations at the nuclear center. $\rho_e(r_0)/\max(\rho_e(r))$ is the charge density at radius $r_0$ relative to its maximum value in the nucleus.

|  | $^4$He | | $^{12}$C | $^{40}$Ca | $^{48}$Ca | $^{56}$Fe | $^{208}$Pb |
|---|---|---|---|---|---|---|---|
| Model | $S$ | $F3$ | $S$ | $F3$ | $F3$ | $F2$ | $F2$ |
| $c$, fm |  | 0.964 |  | 3.766 | 3.7369 | 4.111 | 6.624 |
| $b$, fm |  | 0.322 |  | 0.586 | 0.5245 | 0.558 | 0.549 |
| $w$ |  | 0.517 |  | -0.161 | -0.030 |  |  |
| Ref. | [8] | [8] | [9] | [10] | [11] | [12] | [13] |
| $r_0$, fm | 0.723 | 0.726 | 1.419 | 2.858 | 2.565 | 2.787 | 5.182 |
| $-\Delta r_0$, fm | 0.188 | 0.188 | 0.233 | 0.258 | 0.171 | 0.442 | 0.353 |
| $+\Delta r_0$, fm | 0.127 | 0.128 | 0.185 | 0.237 | 0.155 | 0.354 | 0.323 |
| $\rho_e(r_0)/\max(\rho_e(r))$ | 0.94 | 0.88 | 0.91 | 0.75 | 0.91 | 0.91 | 0.93 |

2.2 The hypothesis that the protons from the nuclear interior do not give an observable contribution to QES is based on the assumption that the properties of intranuclear protons, being at high-density nuclear matter conditions, are likely to be different from the properties of free protons. In this case, the region of existence of modified protons, according to Eq.(3), is $r < r_0$, where the nuclear matter density is $\rho_m(r) > \rho_m(r_0)$. Let P $= \rho_m(r_0)$ be the critical density. It can be treated either as a density corresponding to a jump-like modification of a proton, or as a certain average over a limited range of density values, where the proton properties change significantly. The critical density is the characteristic of proton interaction with the medium. Therefore, the $P$ value cannot depend on the choice of the nucleus, from the study of which it is determined. In other words, if we denote the $P$ value obtained from the data on the nucleus $A$ by $P_A$, then the condition

$$P_A \neq f(A) \text{ or } P = \text{const.} \qquad (4)$$

should be satisfied. If there are several $P_A$ values obtained from the experimental data on different nuclei, then this condition can be considered as a test for the proposed hypothesis.

2.3 The $P_A$ values estimated in [14] are in close agreement between themselves for all the nuclei under consideration: $^{12}$C, $^{48}$Ca, $^{56}$Fe and $^{208}$Pb. The calculations of $P_A$ in the present work are given in Table 3. They cover all the spherical nuclei, for which the experimental values $S' < 1$ are known. It is seen that $P_A$ of $^4$He is appreciably different from $P_A$ of other nuclei. To reconcile this contradiction, let us pay attention to the definition of $P$. It corresponds to the nuclear matter density at the point $r_0$ and, hence, can be considered as a density of matter surrounding the proton of zero radius. However, the real proton has a space length equal to approximately 2 fm, and the nuclear density varies within wide limits over this length in the vicinity of $r = r_0$. In view of this, we average $\rho_m(r)$ over the surface of proton "contact" with the nuclear medium, i.e., over the surface of sphere of radius $R$. The parameter $R$ is an effective radius of nuclear medium–proton interaction and can be considered as a formal parameter. Note that the introduction of this parameter by no means implies the use of the model of uniform matter distribution in the proton. For a spherical nucleus, the expression for the density averaged in this manner takes the form

$$\rho_m(r,R) = \frac{1}{2} \int_{-1}^{1} \rho_m\left(\sqrt{r^2 + R^2 - 2rRx}\right) dx . \tag{5}$$

It is obvious that in consequence of a short-range character of nuclear forces, the maximum $R$ value is limited by the radius of nuclear volume occupied by the proton ($\max(R) \approx (4/3\pi\rho_m)^{-1/3}$, i.e., between 1.13 and 1.19 fm), and its probable (or possible) value is close to the radius of equivalent uniform charge distribution of the proton ($R_{eq} = 1.03$ fm).

The critical density taking into account the averaging (5) is written as

$$P(R) = \rho_m(r_0, R). \tag{6}$$

It is not difficult to note that $P(R=0)$ and $P(R \neq 0)$ differ little from each other until the $r_0$ value becomes close to $R$ (the case of lightest nuclei). The calculated $P(R_{eq})$ values are listed in Table 3. Here we have $P_A(R_{eq})/P_A(0) = 0.95$ for $^{208}$Pb and $P_A(R_{eq})/P_A(0) = 0.65$ for $^4$He. It is of particular importance that now the $P_A(R_{eq})$ values for $^4$He and $^{208}$Pb are equal to within the error.

Let us average $P_A(R_{eq})$ over all six nuclei: $\overline{P}(R_{eq}) = 0.138$. The scatter in $P_A(R_{eq})$ about $\overline{P}(R_{eq})$ is characterized by the average $\chi^2$ per nucleus, $\overline{\chi_i^2} = 1.1$, this corresponding to the test (4).

Table 3. $P_A(R)$ is the value obtained from the data on different nuclei $A$ for $R = 0$, $R_{eq}$, $R_0$ ($R_{eq}=1.03$ fm; $R_0=0.97$ fm). The errors of $P_A(R_0)$ are equal to the errors of $P_A(R_{eq})$, i.e., $\Delta P_A(R_{eq})$.
When calculating the $P_A$ values for the nuclei $^{40,48}$Ca we used the $F3$ model for density distributions of protons (p) and neutrons (n) with the parameters from [15]:
$^{40}$Ca: $c_p$=3.676 fm, $b_p$=0585 fm, $c_n$=3.65 fm, $b_n$=0.60 fm, $w_p$=$w_n$ = −0.102;
$^{48}$Ca: $c_p$=3.744 fm, $b_p$=0526 fm, $c_n$=3.90 fm, $b_n$=0.57 fm, $w_p$=$w_n$ = −0.03.
In other cases, both the distributions were taken, with an accuracy of normalization, the same as for the charge density (see Table 2).

| | $^4$He | | $^{12}$C | $^{40}$Ca | $^{48}$Ca | $^{56}$Fe | $^{208}$Pb |
|---|---|---|---|---|---|---|---|
| Model | $S$ | $F3$ | $S$ | $F3$ | $F3$ | $F2$ | $F2$ |
| $P_A(0)$ fm$^{-3}$ | 0.211 | 0.210 | 0.150 | 0.127 | 0.156 | 0.149 | 0.149 |
| $P_A(R_{eq})$ fm$^{-3}$ | 0.136 | 0.134 | 0.125 | 0.115 | 0.145 | 0.140 | 0.142 |
| $-\Delta P_A(R_{eq})$ fm$^{-3}$ | 0.011 | 0.010 | 0.011 | 0.012 | 0.006 | 0.011 | 0.007 |
| $+\Delta P_A(R_{eq})$ fm$^{-3}$ | 0.014 | 0.014 | 0.010 | 0.012 | 0.006 | 0.013 | 0.010 |
| $P_A(R_0)$ fm$^{-3}$ | 0.143 | | 0.141 | 0.127 | 0.116 | 0.146 | 0.141 | 0.143 |

2.4 Since the calculations with $R = R_{eq}$ have shown agreement with the data, we can regard $R \approx R_{eq}$, and to define the $R$ value more exactly, we take condition (4) as an input requirement. Fig.3 shows the function $P(R)$ for several nuclei. Requirement (4) means close agreement between the values of all the functions investigated at a certain $R_0$ value of the argument $R$.. Analytically, this can by written as a minimum of the following function

$$U(R) = \sum_{i \neq j} \frac{\left|P_i(R) - P_j(R)\right|}{\left[\Delta P_i(R)\right]^2 + \left[\Delta P_j(R)\right]^2} \bigg/ \sum_{i \neq j} \frac{1}{\left[\Delta P_i(R)\right]^2 + \left[\Delta P_j(R)\right]^2} , \tag{7}$$

where $\Delta P(R) = [P(R, r_0 - \Delta r_0) - P(R, r_0 + \Delta r_0)]/2$, and the indices $i$ and $j$ are the numbers of nuclei included in the consideration. Fig.4 represents the function $U(R)$ for various groups of nuclei. It is seen that only the presence of $^4$He in the group provides the existence of the minimum. The position of this minimum does not virtually depend on the exclusion of this or that nucleus, except for $^4$He, from the group and corresponds to $R_0 = 0.97$ fm. This suggests that the error in defining the $R_0$ value is determined by the dependence of $P(R_0)$ of $^4$He on $r$ within $r_0 \pm \Delta r_0$; whence it follows that $\Delta R_0 = \pm 0.13$ fm.

At $R_0 = 0.97$ fm, the mean density for six nuclei is $\overline{P}(R_0) = 0.140$ fm$^{-3}$, and $\overline{\chi_i^2} = 1.1$. The greatest contribution to $\chi^2$ is given by $P_A(R_0)$ of $^{40}$Ca: $\mid \overline{P}(R_0) - P_{40}(R_0) \mid / \Delta P_{40}(R_0) \approx 2$. Not discussing this fact in detail, we note that the deviation of one value from the mean of six data for two standard errors does not contradict to the statistics (variations in one value by two standard errors about the mean with respect to six values are not contrary to statistics). In this case, it is permissible to exclude the drastically different value from the averaging procedure. On this basis we take the mean with respect to five nuclei (i.e., without $^{40}$Ca) to be 0.142 fm$^{-3}$ in this case at $\overline{\chi_i^2} = 0.51$.

The error in $P$ is made up of the errors in the averaged $P_A(R_0)$ and of the error in the calculation of $R_0$. The last error is defined as $\mid \overline{P}(R_0 \pm \Delta R_0) - \overline{P}(R_0) \mid$, it follows from $\overline{P}$ as a function of $R$ shown in Fig.5. So, we have

$$P(R_0) = 0.142 \pm 0.005 \text{ fm}^{-3}.$$

This value of $P(R_0)$ was obtained in [16]. Fig.6 illustrates the difference between the functions $\rho_m(r,R_0)$ and $\rho_m(r)$ in the $^4$He case. It is seen from the figure that the $\rho_m(r,R_0)$ function of $^3$He does not reach the $P(R_0)$ value. The last fact explains $S' = 1$ for the nuclei with $A < 4$.

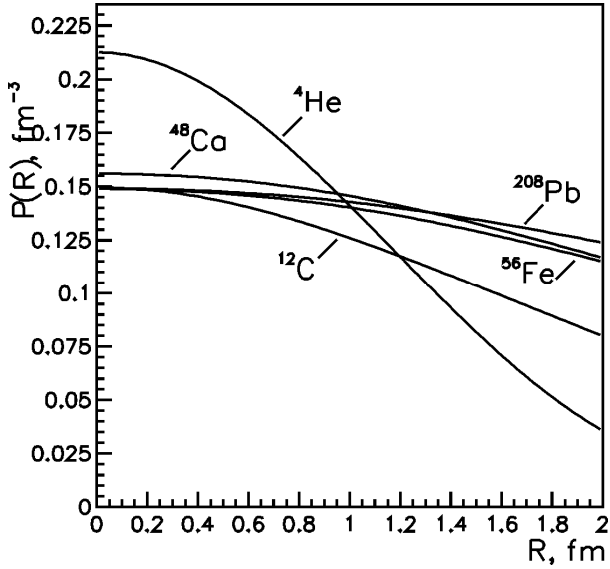

Fig.3. The critical density functions $P(R)$ for the nuclei $^4$He, $^{12}$C, $^{48}$Ca, $^{56}$Fe, $^{208}$Pb.

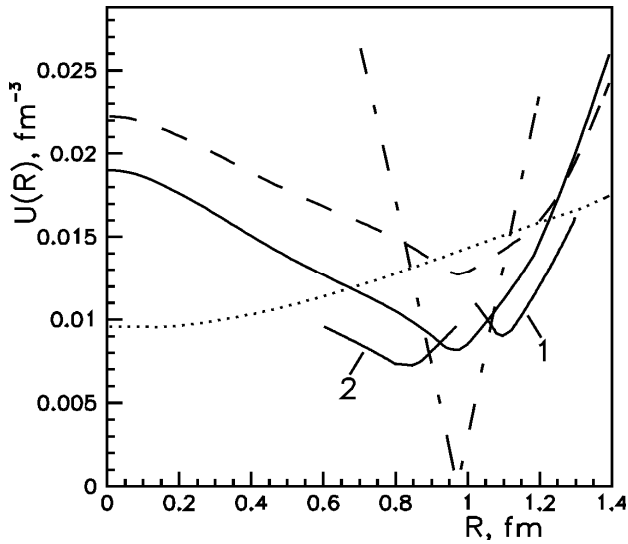

Fig.4. $U(R)$ is the average difference of functions $P(R)$ for different nuclei in the groups:
the solid line is for $^4$He, $^{12}$C, $^{48}$Ca, $^{56}$Fe, $^{208}$Pb
Labels 1 and 2 show the calculations for this group in the region of minimum, where the function $P(R)$ of $^4$He is equal to $\rho_m(r_0 - \Delta r_0, R)$ and $\rho_m(r_0 + \Delta r_0, R)$), respectively;
$^4$He, $^{12}$C, $^{40}$Ca, $^{48}$Ca, $^{56}$Fe, $^{208}$Pb – dashed line;
$^4$He, $^{208}$Pb – dot-and-dash line;
$^{12}$C, $^{40}$Ca, $^{48}$Ca, $^{56}$Fe, $^{208}$Pb - dotted line.

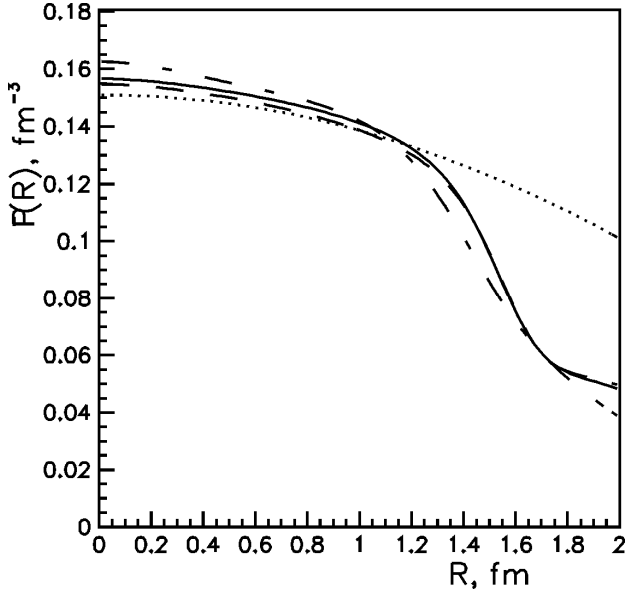

Fig.5. Average critical density $P(R)$ as a function of $R$ for different groups of nuclei. Designation of groups is the same as in Fig.4.

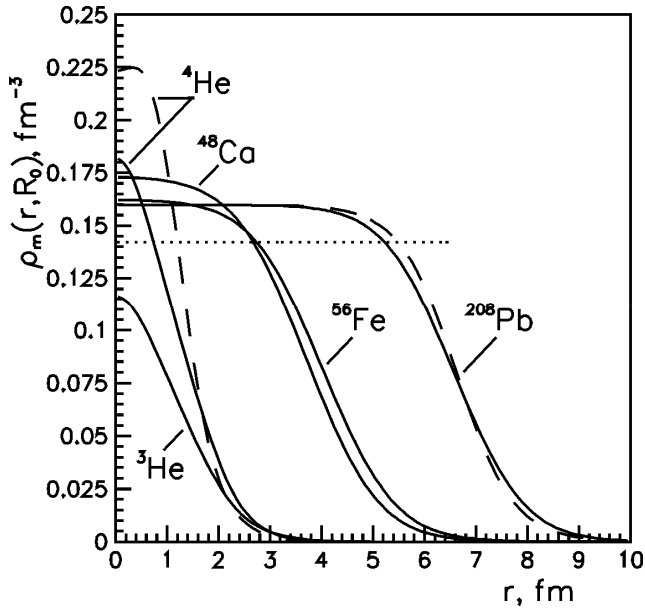

Fig.6. Distribution of averaged nuclear matter density $\rho_m(r, R_0)$ in the nuclei $^3$He, $^4$He, $^{48}$Ca, $^{56}$Fe, $^{208}$Pb. $\rho_m(r, R_0)$ is determined in the text by Eq.(5). The solid line shows the calculation with $R_0 = 0.97$ fm. For comparison, the variant of calculation is given for the nuclei $^4$He and $^{208}$Pb with $\rho_m(r, R=0) = \rho_m(r)$ (dashed line).

The dotted line is $P = 0.142$ fm$^{-3}$.

### 3.Manifestation of modified proton properties in the Coulomb sums

3.1 We can represent the CS as

$$S' = S'^0 + S'^M, \qquad (8)$$

where $S'^0$ is the contribution of nonmodified protons (n.m.p.) determined by expression (3), and $S'^M$ is the contribution of $Z_M$ modified protons (m.p.). It is not difficult to see that the second term of sum (8) can be written as

$$S'^M(q) \cong D\,[g^p_E(Q^2)/\,G^p_E(Q^2)]^2, \qquad (9)$$

where $g^p_E(Q^2)$ is the m.p. electric form factor, $Q^2 = q^2(1-q^2/(4M^2))$. We assume the Gaussian charge density distribution for the proton. Then we have $[G^p_E(Q^2)]^2 = \exp(-Q^2 <r^2>/3)$ and

$$S'^M(q) \cong D\exp(-Q^2\,\Delta\,/3). \qquad (10)$$

Here $\Delta = <r_M^2> - <r^2>$; $<r_M^2>^{1/2}$ and $<r^2>^{1/2}$ are the r.m.s. radii of m.p. and n.m.p., correspondingly. The parameter $\Delta$ is related to $k = <r_M^2>^{1/2}/<r^2>^{1/2}$ by the following expression:

$$k = (1+\Delta/<r^2>)^{1/2}, \qquad (11)$$

$<r^2>^{1/2} = 0.8$ fm. From expressions (8), (9), (10) it follows that with m.p. existent in the nucleus ($<r_M^2> \neq <r^2>$), the CS is the function of $q$ at $q > 2$ fm$^{-1}$, too: $S'(q)$. Using $D$ and expression (10), we write down Eq.(8) as

$$S'(q) \cong 1 - D + D \exp(-Q^2 \Delta /3). \qquad (12)$$

Let us assume that all protons in the nucleus are modified ($D = 1$). Then, according to (12) and (11), and using the averaged values of Table 1 ($q = 2.5$ fm$^{-1}$) we obtain $k = 1.07$, 1.12, 1.21 for $^4$He, $^{48}$Ca, $^{208}$Pb nuclei, respectively.

It was assumed in Section 2 that in $S'$ there is no contribution from m.p. ($S'^M = 0$). As the calculations in this Section, based on both the experimental data and Eq.(3), lead to noncontradictory physical conclusions, it remains to assume that the $S'^M$ value is limited by the measurement error. Proceeding from this and averaging the values of function (10) we derive

$$\frac{D_{max}}{n} \sum_{i=1}^{n} \exp[-Q^2 \Delta /3] \leq \delta S', \qquad (13)$$

or

$$\sum_{i=1}^{n} \exp\left[-q_i^2 \left(1 - \frac{q_i^2}{4M^2}\right)\frac{\Delta}{3}\right] \leq \frac{n \delta S'}{D_{max}}. \qquad (13a)$$

Here $q_i$ are the $q$ values of experimental $S(q,\omega_{max})$, the averaging over $n$ of which gave the $S'$, $D_{max} = D + \delta D$ values. The variables $\delta D$ and $\delta S'$ are determined by the measurement errors.

3.2 The left-hand part of inequality (13a) is maximum at the minimum $\Delta$ value. Therefore, this inequality determines the minimum $<r_M^2>^{1/2}$ value. Since the available data exhibit the smallest error $S'$ for $^{48}$Ca, and the greatest $D_m$ value for $^{208}$Pb, then the data of these nuclei appear most suitable for estimating lower bounds of $<r_M^2>^{1/2}$.

If the data on a certain nucleus are considered irrespective of the results of measurements on other nuclei, then $D_{max} = D + \Delta S'$ and $\delta S' = 2\Delta S'$ (variant 1). However, when solving this problem one can proceed from the assumption that there exists the critical nuclear matter density (section 2), whose value was obtained by using all the known data of other nuclei. In this case, $P_{min} = P(R_0) - \Delta P(R_0)$ corresponds to the lowest $<r_M^2>^{1/2}$ value. According to eq.(3), $D_{max}$ can be written as

$$D_{max} = \frac{4\pi}{A} \int_0^{r_0'} \rho_m(r, R_0)\, r^2 dr, \qquad (14)$$

where $r'_0$ is determined by the equality $\rho_m(r'_0, R_0) = P_{min}$. In this case, $S'^0$ from Eq.(8) will be $S'^0_{min} = 1 - D_{max}$. From here, we have $\delta S' = (S' + \Delta S') - S'^0_{min}$ (variant 2).

The calculation of min($<r_M^2>^{1/2}$) by variants 1 and 2 gives
$k_{min} = 1.40$, 1.38 from $^{48}$Ca data;
$k_{min} = 1.39$, 1.41 from $^{208}$Pb data.

## 4. Concluding comments

4.1 The theoretical calculations give the $<r_M^2>^{1/2}$ value smaller (e.g., $k = 1.14$ and 1.21 in [17] and [12], respectively) than the one obtained here. In the present work, the min($<r_M^2>^{1/2}$) value is determined by the errors attributed to the experimental CS values. This raises the question about the validity of the $\Delta S'$ value. So, if the real $\Delta S'$ value is greater, then the resulting min($<r_M^2>^{1/2}$) value will become smaller.

4.2 It should be noted that one of the important results of the present work is the development of the way to calculate the CS of any nucleus at $q > 2$ fm$^{-1}$ using its functions $\rho_m(r)$ and $\rho_e(r)$. First, from the equation $\rho_m(r_0, R_0) = P(R_0) \pm \Delta P(R_0)$ we calculate $r_0 \pm \Delta r_0$ of the given nucleus, and then, using eq.(3), we can calculate $S' \pm \Delta S'$.

4.3 It has been established that the charge distribution in the proton depends on the density of its surrounding mass; consequently, the proton structure is dependent on the nuclear forces acting on the proton. But then, owing to the charge independence of nuclear interaction, under conditions when the proton changes its properties, the neutron has to change its properties, too. From this it follows that the conclusion on the peculiarity of the intranuclear proton form factor should be interpreted as a manifestation of nucleon modification in the nuclear medium, and all the conclusions on the localization of modified protons in the nucleus should be extended to modified nucleons.

To summarize, in the approach proposed each nucleus with $A \geq 4$ is defined as a structure in the interior of which the modified nucleons are located, and in the outer diffuse layer there are nonmodified nucleons.

4.4 Here we have considered the simplest version of proton modification, which assumes that this process takes place in the limited range of densities of nuclear medium surrounding the proton, viz., from $P(R_0) - \Delta P(R_0)$ to $P(R_0) + \Delta P(R_0)$. It has been found that the known experimental values of CS for the nuclei with $2 \leq A \leq 208$ are not contradictory to this version when $P(R_0) = 0.142$ fm$^{-3}$ and $\Delta P(R_0) = 0.005$ fm$^{-3}$. The fact that $\Delta P(R_0) << P(R_0)$ indicates that the dependence of proton modification on the medium density is likely to have the jump-like character similarly to the behaviour of characteristics during phase transition. The CS-based investigation of more complicated relationships between proton modification and medium density is not efficient, because it calls for more accurate CS data than those available at present.

4.5 It is of interest to know what new information can be provided by the development of experiments on CS measurements. As simple estimates show, measurements of S' with an accuracy 1.5 to 2 times better than that for the now available data or (and) a certain advance of measurements towards $q > 3$ fm$^{-1}$ will enable one to observe the dependence of $S'$ on $q$. Since the function $S^M(q)$ is essentially a normalized (in a certain way) form factor of the modified proton, these measurements will make it possible to find the charge density distribution in this proton and $<r_M^2>^{1/2}$. If the measurements for different nuclei give the same parameters of the m.p. charge distribution, this would give evidence that the proton in this state has always definite and unchanged characteristics. Should it appear that the characteristics of the m.p. depend on the nuclear properties, then these data will serve to investigate the process of modification itself.